\documentclass[aps,pra,reprint,groupedaddress,showpacs]{revtex4-1}
\usepackage{subfigure}
\usepackage{graphicx}
\usepackage{natbib}
\usepackage{amsmath}
\usepackage{amssymb}

\begin{document}

\title{Ultracold atoms in an optical lattice with dynamically variable periodicity}


\author{S. Al-Assam}
\author{R. A. Williams}
\author{C. J. Foot}
\affiliation{Clarendon Laboratory, Department of Physics, University of Oxford, Parks Road, Oxford, OX1 3PU, United Kingdom}

\date{\today}

\begin{abstract}
The use of a dynamic ``accordion'' lattice with ultracold atoms is demonstrated. Ultracold atoms of ${}^{87}$Rb are trapped in a
two-dimensional optical lattice, and the spacing of the lattice is then increased in both directions from 2.2~$\mu$m to
5.5~$\mu$m. Atoms remain bound for expansion times as short as a few milliseconds, and the experimentally measured minimum ramp
time is found to agree well with numerical calculations. This technique allows an experiment such as quantum simulations to be
performed with a lattice spacing smaller than the resolution limit of the imaging system, while allowing imaging of the atoms at
individual lattice sites by subsequent expansion of the optical lattice.
\end{abstract}


\maketitle

Optical lattices create clean, tunable and flexible periodic potentials for ultracold atoms that are an important tool for
investigating the quantum behavior of strongly correlated many-body systems
\cite{PhysRevLett.81.3108,lewenstein2007ultracold,RevModPhys.80.885}. Such investigations require lattices with sub-micron
spacing to ensure that the quantum dynamics occurs on millisecond timescales (and so is not subject to decoherence). The small
spacing also ensures that the tunneling energy $J$ and on-site interaction energy $U$ are comparable ($J/U\sim1$) at a lattice
depth for which the band gap $E_g\gg k_BT$, assuming a typical temperature of tens of nanokelvin. The sub-micron lattice spacing
makes it challenging to observe atoms at individual lattice sites with light or near infrared radiation, and typically
time-of-flight expansion has been used to probe the momentum of the atoms \cite{greiner2002quantum}. Being able to detect the
position of the atoms \emph{in situ} can yield further information about the system, and the ability to address atoms at single
lattice sites is crucial for quantum information processing \cite{jaksch2004optical}. \emph{In situ} imaging of the atoms has
been demonstrated for lattice spacings from 2~$\mu$m upwards \cite{NatPhysWeiss,ItahDirect2010,Weitzaddressing}, however at these
spacings the single-atom tunneling is negligible on typical experimental timescales. Scanning electron microscopy has been used
to detect single atoms in a 0.6~$\mu$m optical lattice and determine the density distribution by summing over multiple images
\cite{Ott2008electron}. Recently there has been significant progress on directly imaging single atoms trapped in optical lattice
sites at sub-micron lattice scales using fluorescence imaging \cite{ScienceGreiner,Blochsingle}: in both cases sophisticated
optical arrangements were used. The difficulty of obtaining the theoretical maximum resolution in an optical imaging system
increases dramatically as the numerical aperture increases.

Our method of imaging such systems is to use a lattice with a dynamically variable spacing. Ultracold atoms are prepared in a
lattice with small spacing and then the lattice potential is expanded to facilitate imaging.  This allows individual lattice
sites to be resolved without the need for a complex or expensive microscope arrangement. This approach could also be used for
addressing individual lattice sites; a previous experiment \cite{Weitzaddressing} demonstrated addressing single lattice sites
spaced by 5.3~$\mu$m. Increasing the spacing between the trapped atoms also reduces the error in qubit readout fidelity
introduced by cross talk between neighboring atoms \cite{PhysRevA.81.040302}. Dynamically varying the lattice spacing could
provide a promising technique for measurement based quantum computing where very high ground state fidelities are required
--- after a cluster state is created \cite{Briegel2001QC,VaucherJaksch} the optical lattice could be expanded to allow
measurements to be carried out.

Diffraction from a 1D lattice with variable periodicity \cite{Huckans}, and investigation of the dynamics of a Bose-Einstein
condensate (BEC) loaded into a large period 1D lattice with tuneable spacing \cite{fallani2005bose} have been demonstrated;
neither of these arrangements, however, were capable of changing the lattice spacing whilst keeping the atoms trapped. An
arrangement for generating 1D dynamically varying accordion lattices has been shown \cite{li2008real} in which mechanical means
were used to vary the lattice spacing from 0.96~$\mu$m to 11.2~$\mu$m over one second (although it was not demonstrated with
trapped ultracold atoms).

Here we demonstrate the first use of an accordion lattice arrangement to dynamically increase the optical lattice spacing while
the atoms are trapped. Unlike in the work described above \cite{Huckans,fallani2005bose,li2008real}, the lattice spacing is
increased in two dimensions, and the arrangement allows the optical potential to be varied smoothly and quickly, free of
mechanical vibrations. We show that the atoms remain trapped in the optical lattice for ramp times as short as a few
milliseconds, well within the timescales of a typical BEC experiment.

The accordion lattice is generated using dual axis acousto-optic deflectors (AODs), and an optical arrangement described in
detail in \cite{williams2008dynamic}. Briefly, the optical lattice is formed in the focal plane of a multi-element objective lens
made according to the design described in \cite{Alt2002142}. The intersection angle of the beams determines the lattice spacing
$d$, i.e.\ $d$ = $\lambda F/D$, where $\lambda$ is the wavelength and $F$ is the focal length of the lens. $D$ is the distance
between the beams in the back focal plane of the lens, which is controlled by the deflection angle from an AOD. Thus the lattice
spacing can be varied dynamically by changing the frequency of the radio frequency radiation controlling an AOD. This arrangement
provides an extremely flexible optical lattice which can be rotated \cite{williams2010observation}, or expanded and contracted as
described here.  The minimum lattice spacing achievable is given by the numerical aperture of the lens, which in our case is
0.27. The optical lattice is formed by the interference of near infrared radiation at 830~nm provided by a Ti:sapphire laser
(Coherent MBR-110). The optical lattice potential has the form
\begin{equation} \label{eqn:olpot}
V(x,y,t) = V_l(x,y)\{\cos^2[\pi x/d(t)]+\cos^2[\pi y/d(t)]\};
\end{equation}
the spatially-dependent lattice depth $V_l(x,y)$ has a Gaussian distribution $-V_0\exp[-2(x^2+y^2)/w^2]$, where $V_0$ is the peak
lattice depth and $w$ is the waist.

\begin{figure}[h]
\begin{center}
  \subfigure[~Before ramp]{\label{fig:beforeramp}\includegraphics{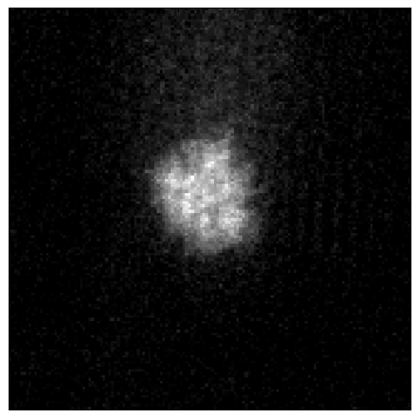}}
  \subfigure[~After ramp]{\label{fig:afterramp}\includegraphics{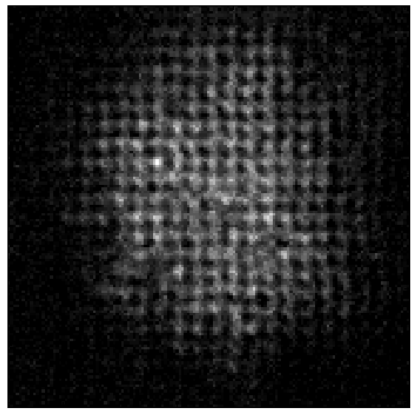}} \\
  \subfigure[~Before ramp]{\label{fig:beforeramp2}\includegraphics{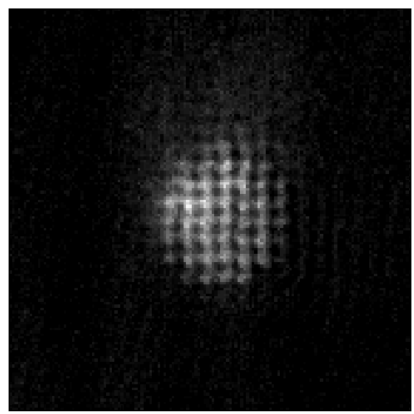}}
  \subfigure[~After ramp]{\label{fig:afterramp2}\includegraphics{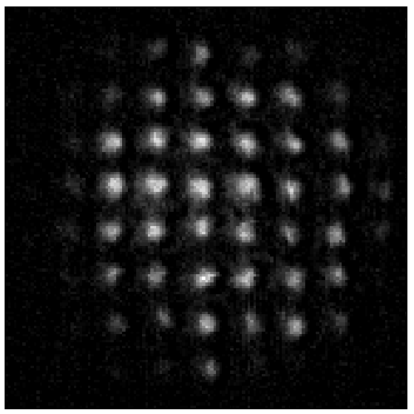}} \\
  \caption{Absorption image of atoms trapped in a two-dimensional optical lattice before and after the lattice spacing has been ramped by: a factor of 2.5 from 2.2~$\mu$m to 5.5~$\mu$m in 10~ms [(a) and (b)]; a factor of 2.5 from 5.5~$\mu$m to 13.75~$\mu$m in 50~ms [(c) and (d)]. The field of view is 120~$\mu$m by 120~$\mu$m in each image.}
\end{center}
\end{figure}

We start with a nearly pure BEC of 80,000 ${}^{87}$Rb atoms in the $F=1,m_{F} = -1$ state in a combined magnetic and optical trap
with harmonic trapping frequencies $\{\omega_r,\omega_z\} = 2\pi \times \{20.1,112\}~\mathrm{Hz}$. This trap is on for the
remainder of the experimental sequence. An optical lattice potential of the form given in Eq.\ (\ref{eqn:olpot}) with $V_0 = h
\times 1.5$~kHz, $w = 69~\mu$m and $d = 2.2~\mu$m is ramped on in 500~ms. This timescale is long enough to ensure the system is
in the ground state of the optical lattice. Figure \ref{fig:beforeramp} shows an \emph{in situ} absorption image of atoms trapped
in the optical lattice; we first optically pump the atoms to the $F = 2$ hyperfine level, and then probe the atoms by applying a
20~$\mu$s pulse of circularly polarized light resonant with the $5S_{1/2}~F = 2$ -- $5P_{3/2}~F' = 3$ transition. The optical
pumping light is detuned to ensure only a fraction of the atoms are pumped to the $F=2$ level, reducing the optical density in
the lattice to order one to prevent saturation of the absorption images. The individual lattice sites cannot be resolved, because
the finite axial extent of the cloud, which is 11~$\mu$m diameter or more, is comparable with the depth of field of the imaging
system. The trapped cloud has radius $r_0 = 18~\mu$m, which indicates that there are around 8 lattice sites filled in each
direction, with around 1200 atoms in each central lattice site.

The lattice depth is then ramped to $h \times 37 $~kHz in 5~ms, which cuts off tunneling between the lattice sites. This ensures
that the lattice dynamics are frozen during the expansion of the lattice. The ramp time is short compared to the single-particle
tunneling time, however it is long enough to ensure that the atoms remain in the lowest energy band of the optical lattice, i.e.\
$\langle\partial H/\partial t \rangle\ll E_{g}^2/\hbar$.

The lattice spacing is varied as a smoothed ramp given by the Gauss error function $d(t) = d_0\textbf{(}1 +
(S-1)\{\mathrm{erf}[7.14(t/t_r-0.5)]+1\}/2\textbf{)}$, where $d_0$ is the initial spacing (2.2~$\mu$m), $S$ is the ratio of the
final spacing to the initial spacing and $t_r$ is the ramp time (which is varied). The peak lattice depth $V_0$ is kept constant
during the ramp, and $S$ is 2.5. In principle the lattice spacing can be expanded to any size provided that the lattice spacing
remains smaller than the beam waist $w$. In practice, to ensure that atoms at the edge of the cloud are not lost, $S \lesssim
w/r_0$ should be satisfied. For our parameters, after a 2.5 times expansion, the radius is 45~$\mu$m, and the depth at the
outermost lattice site has dropped to about 40\% of the depth at the center of the lattice. The final lattice spacing,
5.5~$\mu$m, is large enough to be resolved by our imaging system, as shown in Fig.\ \ref{fig:afterramp}. The number of lattice
sites filled is consistent with the dimensions of the cloud before expansion. To confirm that the accordion lattice is working as
expected, we also expand by a factor of 2.5 from an initial lattice spacing that is large enough to be resolved before expansion
(5.5~$\mu$m to 13.75~$\mu$m). The result, presented in Fig.\ \ref{fig:beforeramp2} and \ref{fig:afterramp2}, shows that the atoms
do remain bound for this ramp.

During the expansion of the lattice, the central lattice site remains stationary \cite{Footnote1}, while the first lattice site
moves with speed $v_{1} = \dot{d}$, the second lattice site moves with speed $v_{2} = 2\dot{d}$, and the $n^{\mathrm{th}}$
lattice site moves with speed $v_{n} = n\dot{d}$. To determine the evolution of the system during the ramp in lattice spacing, we
describe the atoms at each lattice site by localized wave functions $|\psi_{n_x,n_y}\rangle$, evolving according to
\begin{equation} \label{eqn:labframe}
i\hbar\frac{\partial|\psi_{n_x,n_y} (t)\rangle}{\partial t} = \left(\frac{p^2}{2M} + V(x,y,t)\right)|\psi_{n_x,n_y} (t)\rangle,
\end{equation}
where $M$ is the mass of an atom, and interactions are ignored.

If the accordion lattice is to be used for quantum computing, it is important that the ground state fidelity remains high during
the lattice expansion. However, if  the lattice is expanded for the purpose of imaging the atoms, moving the atoms slowly enough
to retain high ground state fidelity is not necessary --- instead there is the less stringent requirement that an atom remains in
a bound state of the same lattice site.

\begin{figure}[h]
\begin{center}
  \includegraphics{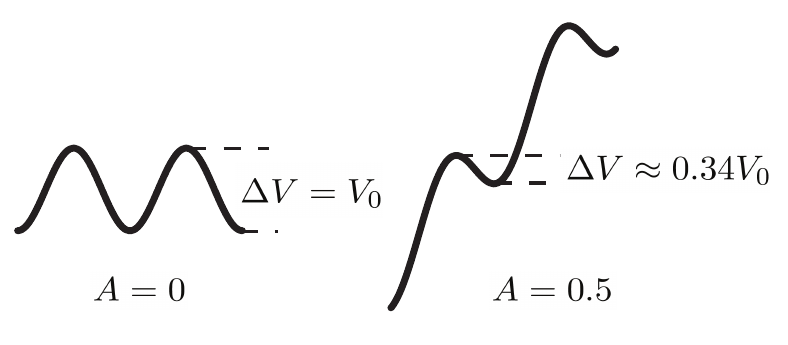}
  \caption{Effective trapping potential at the $n^{\mathrm{th}}$ lattice site in a reference frame accelerating at $n\ddot{d}$, in terms of the scaled acceleration $A = Mn\ddot{d}d/\pi V_0$.} \label{fig:sinpots}
\end{center}
\end{figure}

To estimate the maximum acceleration for which the atoms in the $n^{\mathrm{th}}$ well remain bound, we can move to an
accelerating reference frame in which the  $n^{\mathrm{th}}$ lattice site is at rest. In this reference frame, the lattice
potential is tilted due to a linear potential term $Mn\ddot{d}x$ \cite{Peik1997Bloch} which has the effect of reducing the
potential barrier between neighboring lattice sites, as illustrated in Fig.\ \ref{fig:sinpots}. We would expect atoms to be lost
when the minimum barrier height $\Delta V$ approaches the energy of the trapped atoms, which for our experimental parameters
would occur for accelerations corresponding to a ramp time of around 3 ms. $\Delta V$ decreases with increasing scaled
acceleration $A = Mn\ddot{d}d/\pi V_0$, and so the maximum acceleration can be larger for a deeper lattice, or for a smaller
lattice spacing. This approach gives an order of magnitude estimate for the minimum expansion time, however it does not take into
account the reduction of the barrier height during expansion due to the envelope of the lattice beams, the effect of the magnetic
trap, nor does it take into account excitations to higher bands during the ramp.

We perform a more detailed analysis of the minimum expansion time by numerically solving Eq.\ (\ref{eqn:labframe}). The optical
lattice potential $V(x,y,t)$ has the form given in Eq.\ (\ref{eqn:olpot}), with our experimental parameters ($V_0 = h \times
37$~kHz, $d_0 = 2.2~\mu$m, $w = 69~\mu$m, $S = 2.5$) and the harmonic trapping potential is included. We take the initial wave
function $|\psi_{n_x,n_y} (t=0)\rangle$ to be the ground state of a single well of the sinusoidal potential centered on $x =
n_xd_0$, $y = n_yd_0$. A $6^\mathrm{th}$ order finite difference formula is used to express the second derivative, and the wave
function is evolved using the propagator $U(t) = \exp[-iH(t)\Delta t/\hbar]$ for each time-step $\Delta t$. We solve Eq.\
(\ref{eqn:labframe}) in one dimension along the direction of motion for each lattice site \cite{Footnote2}.

\begin{figure}[h]
\begin{center}
  \subfigure[]{\includegraphics{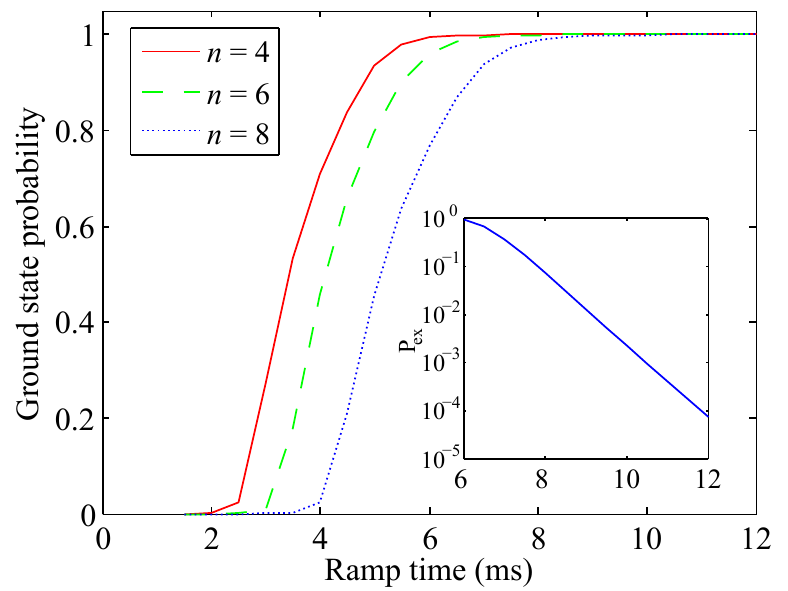}\label{fig:gs}} \\
  \subfigure[]{\includegraphics{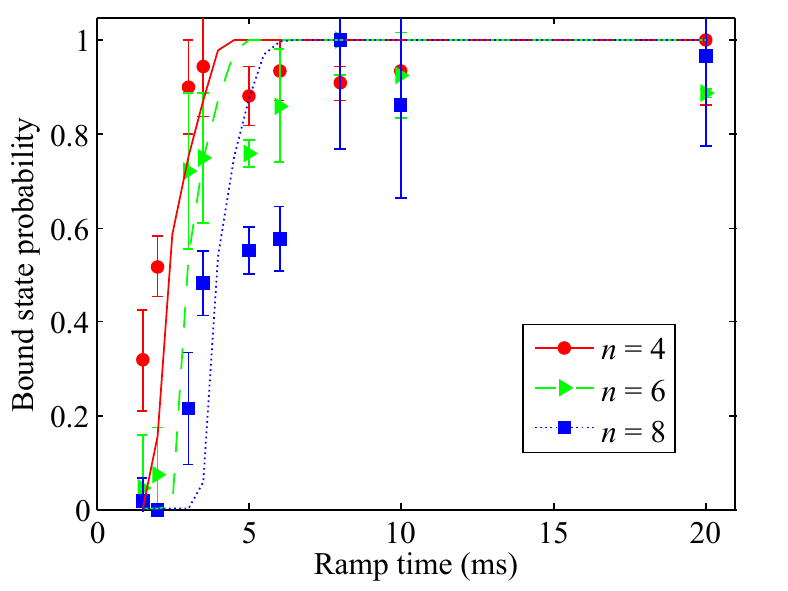}\label{fig:bc}}
  \caption{(color online) (a) Numerically calculated probability of finding a single particle in the ground state of the $\cos^2(x)$ potential well of the $n^{\mathrm{th}}$ lattice site after the spacing is increased from 2.2~$\mu$m to 5.5~$\mu$m for different ramp times. $V_0 = h
\times 37$~kHz. The inset in (a) shows the excited state probability against ramp time for the outermost lattice site ($n_x = 8$,
$n_y = 8$). (b) Probability of finding an atom in a bound state for the same parameters. The lines denote the numerically
calculated probability and the points show experimental data.}
\end{center}
\end{figure}

The probability of an atom remaining in the ground state, $| \langle \psi_{n_x,n_y} (t=t_r)| \psi_{\mathrm{ground}} \rangle |
^2$, is shown in Fig.\ \ref{fig:gs} (we find the probability for the $n^{\mathrm{th}}$ lattice site by averaging over $n_y$ for
each value of $n_x$). For the outermost lattice site ($n_x = 8$, $n_y = 8$) the ground state fidelity is more than $99.99\%$ for
ramps times longer than 12~ms. The time taken to spontaneously scatter a photon is 7~s in our lattice (with $V_{0} = h \times
37$~kHz), and can be several minutes in optical lattices with blue frequency detuning. Thus many expansion and addressing
operations could be carried out with high fidelity before coherence is destroyed due to spontaneous emission.

The probability of an atom remaining in a bound state of the optical lattice is also numerically calculated and we compare this
with experimental results. In the experiment, we expand the lattice as described above for different ramp times. To determine the
probability of an atom in a given site remaining bound, we first sum over the columns of the lattice, then calculate the number
of atoms in site $n$. The number of atoms in the outermost site does not change (within the margin of error) for expansion times
between 8 and 20~ms, so the number of atoms is normalized to the number of atoms in site $n$ for these ramp times. The results
are shown in Fig.\ \ref{fig:bc}.

The experimental and numerically calculated results agree well.  The small discrepancy (the ramp time for a given probability is
slightly longer in the experiment) may be due to the fact that the numerical results are for a single particle only. In the
experiment, however, there are on average a few hundred atoms per lattice site, and the calculations do not take into account the
energy due to repulsive interactions. The close agreement with the numerically calculated results indicates that the main source
of atom losses is the speed of expansion rather than from a source of instability in the ramp, confirming that the ramp of the
optical potential is smooth.

Note that the minimum ramp time has been calculated for only one lattice depth and one form of ramp. With a deeper lattice, the
expansion timescales would be even faster. Also, it is possible that the minimum ramp time could be reduced further by optimizing
the form of the ramp.

In this experiment the initial lattice spacing was 2.2~$\mu$m, and this technique can be easily used to expand from smaller
lattice sizes if a higher numerical aperture (NA) lens is used to form the optical lattice --- the minimum lattice spacing
$d_{\mathrm{min}} = \lambda_l/2\mathrm{NA}$, where $\lambda_l$ is the wavelength of light forming the optical lattice. Note that
$d_{\mathrm{min}}$ is not equal to the imaging resolution; atoms can just be resolved when they are spaced by more than
$d_{\mathrm{res}} = 0.6\lambda_{i}/\mathrm{NA}$, where $\lambda_{i}$ is the wavelength of imaging light. For
$\lambda_l<\lambda_i$ (blue frequency detuning) the lattice spacing is smaller than the imaging resolution.

In our experiments on the accordion lattice we have concentrated on proving its usefulness for expansion of cold atoms in a
periodic potential which is equivalent to increasing the magnification of the optical imaging system, however there are other
modes of operation of this new apparatus: (1) Two, or more, optical lattices of different spacing can be created at the same time
by applying the requisite driving frequencies to the AODs. The light intensities and resulting optical potentials add together
forming a superlattice. In future work cold atoms could be loaded into an optical lattice of double wells by loading atoms into a
lattice of period $d$, expanding the spacing of the wells to 2$d$ and gradually switching on another lattice of spacing $d$ to
split each potential well in two; (2) the technique can be extended to three dimensions by applying an additional accordion
lattice in the axial direction, enabling planes of atoms to be moved out of focus of the imaging system, thus facilitating
imaging of a single plane; (3) the optical lattice can be expanded to decrease the tunneling energy without changing the lattice
depth, allowing investigation of different quantum phases.

The use of an accordion lattice for expanding the period of an optical lattice has been demonstrated for the case of absorption
imaging, but it could also be utilized for single atom fluorescence imaging. The arrangement used to generate the optical lattice
is extremely flexible and allows smooth variation of the optical potential as shown here, and demonstrated previously with a
rotating optical lattice \cite{williams2010observation}.

We thank E. Nugent, S. R. Clark and D. Jaksch for a critical reading of the manuscript. This research is supported by EPSRC
(EP/E041612/1), QIP IRC (GR/S82176/01), and ESF.


\begin{thebibliography}{22}
\bibitem {Footnote1}%
  \BibitemOpen
  We assume that the position of central lattice site corresponds to the center of expansion. In practice the position of the central lattice site will depend on the relative phase of the lattice beams\BibitemShut {NoStop}%
\bibitem {Footnote2}%
  \BibitemOpen
  We have also calculated the ground state probability for one set of parameters in two dimensions, to verify that ignoring the dynamics along the direction perpendicular to motion does not introduce significant errors\BibitemShut {NoStop}%
\end{thebibliography}
\end{document}